# Online Label Recovery for Deep Learning-based Communication through Error Correcting Codes

Stefan Schibisch*, Sebastian Cammerer*, Sebastian Dörner*, Jakob Hoydis† and Stephan ten Brink*
* Institute of Telecommunications, Pfaffenwaldring 47, University of Stuttgart, 70659 Stuttgart, Germany
†Nokia Bell Labs, Route de Villejust, 91620 Nozay, France

***Abstract*—We demonstrate that error correcting codes (ECCs) can be used to construct a labeled data set for finetuning of "trainable" communication systems without sacrificing resources for the transmission of known symbols. This enables adaptive systems, which can be trained on-the-fly to compensate for slow fluctuations in channel conditions or varying hardware impairments. We examine the influence of corrupted training data and show that it is crucial to train based on correct labels. The proposed method can be applied to fully end-to-end trained communication systems (autoencoders) as well as systems with only some trainable components. This is exemplified by extending a conventional OFDM system with a trainable pre-equalizer neural network (NN) that can be optimized at run time.**

## I. Introduction

Attracted by the conceptual simplicity of deep learning (DL) and the huge success in many fields of application, many papers have recently proposed and analyzed ways to benefit from neural networks (NNs) in communications. Therefore, almost any conventional signal processing block has been replaced individually by NN-blocks such as equalization [1], [2], channel coding [3], [4], detection [5], or multiple-input multiple-output (MIMO)-detection [6] up to DL enhanced iterative decoding algorithms [7]. However, with the introduction of end-to-end learning autoencoder systems [8], [9], the system design paradigms change from optimizing individual sub-components with specific task towards systems that *learn to communicate* without the need of any conventional signal processing block.

In *classical* communication system design, it is assumed that most of the physical effects a transmitted signal suffers from are well known and can be conventionally corrected one by one due to an approximated channel/system model. However, minor effects such as temperature changes, non-linearities and tolerance ranges in hardware components, mostly caused by hardware insufficiencies, or other unusual channel parameters (e.g., extreme weather conditions) often cannot be fully considered in those models and, thus, are not optimally compensated. As an alternative approach, NN-enhanced communication promises that the real physical channel with all its known and unknown parameters can be formulated as a *hard-to-describe* black-box. A neural network-based receiver (autoencoder systems [8], where transmitter and receiver are learned together) can be trained such that it compensates all effects of this black-box without requiring a detailed description or model. This is one promising advantage of deep neural networks in communications.

This work has been supported by DFG, Germany, under grant BR 3205/6-1.

A NN-based system may tolerate a certain mismatch of the channel parameters, but the system itself (without re-training) is static as the weights of the NN do not change once deployed. However, adaptivity towards fluctuations of the channel condition of NN-based systems can be achieved by two different approaches:

1) *Include fluctuations in the training*, i.e., the NN is trained to handle these fluctuations and, thus, the system is designed to be adaptive with respect to certain effects. Assuming the effects occur within a certain maximum range (as design parameter of the system).
2) *Finetune* on-the-fly whenever the channel changes and, thus, only consider the effects when they really occur.

Unfortunately, the first approach requires to consider all possible effects during the training (i.e., during the *design* of the system) and, thus, is often not practical.

In this work, we focus on the second approach, i.e., we investigate the possibility of finetuning the system on-the-fly. As already mentioned before, the principal benefits of this approach are adaptivity with respect to effects *not* considered during the design, and initial training simplicity as not all possible effects of the future system need to be considered during the design. However, to ensure that the receiver neural network can handle channel alterations over time it requires to perform a finetuning step periodically, i.e., the receiver has to update its weights during run time. This can be simply done by the stochastic gradient descent (SGD) algorithm (see [10]), but also leads to a fundamental practical problem of NN-based communication systems: *How can the receiver update its weights without knowing what was originally transmitted*, i.e., without having a labeled data set?

This may be solved by a periodically transmitted pilot sequence (or explicitly triggered by the receiver if a feedback-channel is available). But this causes overhead and, at least for the case where no feedback-channel is available and the system changes are very sporadic, this approach is infeasible. As SGD based training typically requires a large amount of training samples, pilot based data acquisition causes a high transmission overhead and needs a suitable protocol to trigger the training data acquisition phase between transmitter and receiver. Both of these problems disappear as soon as the labels are recovered from the error correcting code (ECC) assuming a suitable code. Thus, the main contribution of this work is to embed an ECC into NN-enhanced communication systems such that a labeled training set can be generated on-the-fly without any piloting overhead.

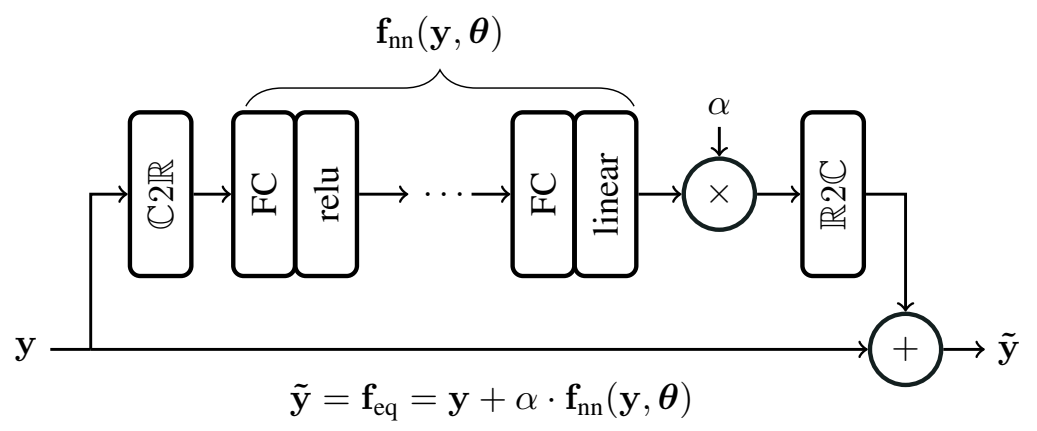

Fig. 1: Pre-equalizer residual NN structure.

We want to clarify that in the context of this work, finetuning always relates to the receiver and, thus, not all transmitter impairments can be compensated as the information may already be lost during the transmission (e.g., strong clipping at transmitter side cannot be recovered at the receiver side). Further, we want to emphasize that the effects we consider in this work are assumed to be slowly changing compared to the run time of the weight update (typically, finetuning requires just a few SGD iterations).

## II. Deep Learning-based Communication

### A. Neural Networks

In this section, we want to introduce relevant notations and provide a brief introduction to deep learning fundamentals. The interested reader is referred to [10] for a general introduction to the field of DL. Most of the tools and NN structures used throughout this work are very basic and well established within the field of deep learning. In its basic form, a feed-forward NN is a directed computation graph consisting of multiple layers of neurons with connections only to neurons of the subsequent layer. Fully connected (FC) layers, in which every neuron of a layer is connected to all neurons of the previous layer, are called dense layers, because of their densely populated weight matrix, and are the structure of our choice for the later discussed NN. Each neuron within such a dense layer sums up all weighted inputs and optionally applies a non-linear activation function, e.g., the rectified linear unit (ReLU) function

$$g_{\text{ReLU}}(x) = \max\{0, x\}$$

before forwarding the output to all connected neurons of the next layer.

Let a dense layer $i$ have $n_i$ inputs and $m_i$ outputs, then it performs the mapping $\mathbf{f}^{(i)} : \mathbb{R}^{n_i} \to \mathbb{R}^{m_i}$ defined by the weights and biases as parameters $\boldsymbol{\theta}_i$. Consecutively applying this mapping from input $\boldsymbol{v}$ of the first layer to the output $\boldsymbol{w}$ of the last layer, leads to the function

$$\boldsymbol{w} = \mathbf{f}\left(\boldsymbol{v}; \boldsymbol{\theta}\right) = \mathbf{f}^{(L-1)}\left(\mathbf{f}^{(L-2)}\left(\ldots\left(\mathbf{f}^{(0)}\left(\boldsymbol{v}\right)\right)\right)\right) \quad (1)$$

where $\boldsymbol{\theta}$ is the set of parameters and $L$ defines the *depth* of the net, i.e., the total number of layers.

TABLE I: Parameters used for the pre-EQ

| Parameter | Value |
|---|---|
| Optimizer | SGD with Adam |
| Learning rate $L_r$ | 0.001 |
| Training SNR $E_b/N_0$ | 10 dB and 14 dB |
| Size of Dense Layers | 256 neurons |
| Number of Dense Layers: | 3 |

*Training* of the NN describes the task of finding suitable weights $\boldsymbol{\theta}$ for a given data set and its corresponding labels (desired output of the NN, *supervised* training) such that a given loss function is minimized. This can be efficiently done with the SGD algorithm [10] in combination with weight-backpropagation as implemented in many state-of-the-art software libraries. We use the Tensorflow library[1]. While all of the calculations within the neural layers are real-valued operations, we implemented all operations within the transmission channel as complex operations, since we are dealing with complex signals. Therefore, all complex-valued signals at the NN input are reshaped to real-valued signals with their real and imaginary part consecutively, as well as all real-valued signals at the NN output are transformed back to complex-valued signals.

### B. Trainable Communication

In principle, receiver finetuning can be applied as soon as the receiver contains trainable components either as separate entity or embedded in the underlying algorithm. A simple but rather universal approach is a *pre-equalizer*[2] NN (see Fig. 1) directly located at the receiver input as shown in Fig. 2. The only necessary condition is the availability of the gradient of all operations in the receiver. Therefore, we use a residual [11] structure as shown in Fig. 1

$$\tilde{\mathbf{y}} = \mathbf{f}_{\text{eq}}(\mathbf{y}) = \mathbf{y} + \alpha \cdot \mathbf{f}_{\text{nn}}(\mathbf{y}, \boldsymbol{\theta})$$

where the scalar $\alpha$ can be optionally an additional trainable parameter. This structure is typically used to overcome the *vanishing gradient* problem, however, in our context the main motivation is that an initialization of $\alpha = 0$ deactivates the NN path, i.e., without any training the receiver behaves exactly as originally designed.

To not restrict us to autoencoder-based communication systems [12] and to demonstrate the conceptual simplicity, we embed a trainable pre-equalizer NN into the receiver of a conventional orthogonal frequency division multiplexing (OFDM) baseline. Thus, we extend a conventional OFDM system by the proposed equalizer (Fig. 2), where the pre-equalization takes place in the time domain. The OFDM system uses $w_{\text{FFT}} = 64$ subcarriers with cyclic prefix (CP) of length $\ell_{\text{CP}} = 8$, quadrature phase-shift keying (QPSK) modulation per subcarrier and minimum mean squared error (MMSE) equalization per subcarrier, for details we refer to the baseline in [13]. Further, details on the pre-EQ NN are provided in Table. I.

[1]www.tensorflow.org

[2]Strictly speaking this is not just an equalizer as the NN can learn any non-linear function.

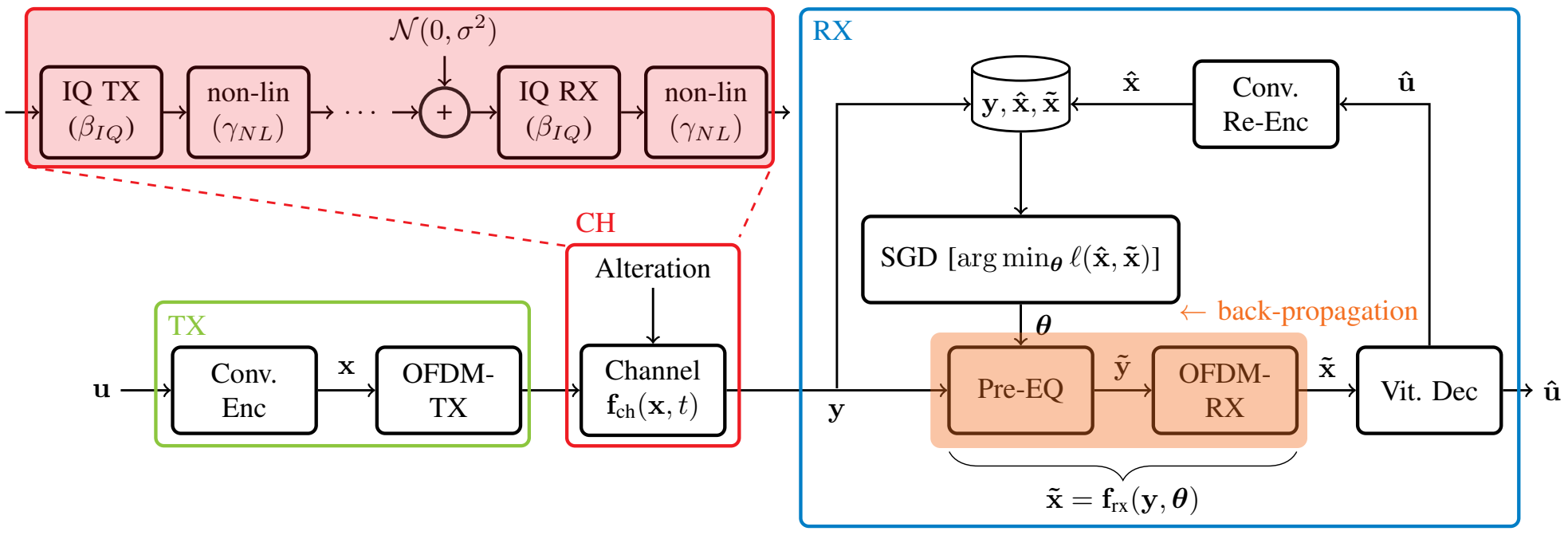

Fig. 2: Communications system including ECC for label recovery.

We want to emphasize that the proposed training method can be applied to autoencoder-based systems (as used in [9], [13]) straightforwardly, but for the sake of comparison and having a broad field of readers in mind, we provide results only for a conventional OFDM system with QPSK modulation.

*1) Demapping and missing gradients:* Especially the demapping (i.e., the assignment from symbols to a binary decision) can be tricky to implement such that a gradient can be calculated for all layers as the labeling is generally defined by a look-up table (LUT) which does inherently not provide any gradient. However, this issue can be solved by changing the classification task into a regression problem. Instead of the cross-entropy loss between the estimated bit sequence and the original sequence, the training can also optimize the mean squared error (MSE) loss between the transmitted symbol (requires OFDM transmitter block at receiver side) and the estimated symbol at the receiver before demapping. This makes the training procedure independent of the underlying labeling. For further details on the loss functions we refer to [12], [4], [10].

*2) Initial training of the pre-equalizer NN:* In case of a fully NN-based receiver, like the autoencoder system [12], the NN is already optimally trained for all considered fluctuations per design, as described in Section 1. Such a system directly improves on labeled finetuning training of the receiver part, because the weights only need slight updates. However, the considered system (Fig. 2) only has a trainable pre-equalizer. Thus, it takes some training time for the pre-equalizer NN to actually improve the receiver's decisions, which we refer to as *initial training phase*. During this initial training phase, the scalar $\alpha$ is slowly increased by the training process and the pre-equalizer NN gains more influence with more training confidence. To skip this transient training part for our simulations, we initially trained the pre-equalizer NN for a specific range of shortcoming factors with perfect label knowledge and only the finetuning training updates are based on recovered labels.

## III. Online Label Recovery through ECC

The proposed system, as depicted in Fig. 2, consists of a conventional OFDM transmitter and an OFDM receiver extended by a trainable pre-equalizer (see Sec. II-B). We use a terminated non-systematic convolutional (NSC) code with encoding memory $\mu = 2$, polynomials 05 and 07 and, thus, of rate $r = 0.5$.[3] The encoder maps $k$ information bits $\mathbf{u}$ into a codeword $\mathbf{x}$ of length $N$. After passing through the channel $\mathbf{f}_{ch}(\mathbf{x}, t)$, the (noisy) sequence $\mathbf{y}$ can be observed at the receiver (consisting of the pre-equalizer and the conventional OFDM receiver), which then provides an estimate $\tilde{\mathbf{x}}$ on the transmitted codeword. The decoder uses the Viterbi algorithm with traceback length $5(\mu + 1)$ and outputs the corrected information bit sequence $\hat{\mathbf{u}}$. A final re-encoding step provides the (estimated) codeword $\hat{\mathbf{x}}$.

During data transmission, the receiver collects a training set consisting of observations of the received sequence $\mathbf{y}$, the estimated codeword $\tilde{\mathbf{x}}$ and corrected codeword $\hat{\mathbf{x}}$ as label for the training.[4] For a given loss function $\ell(\mathbf{x}, \mathbf{y})$ the task of the SGD is then defined as

$$\boldsymbol{\theta}_{\text{new}} = \arg\min_{\boldsymbol{\theta}} \ell(\hat{\mathbf{x}}, \tilde{\mathbf{x}}) = \arg\min_{\boldsymbol{\theta}} \ell(\hat{\mathbf{x}}, \mathbf{f}_{\text{rx}}(\mathbf{y}, \boldsymbol{\theta}_{\text{old}})).$$

The finetuning can be either triggered periodically or whenever a certain bit error rate (BER) threshold is exceeded, this threshold has to be chosen such that the ECC can recover (almost) all errors in the codeword. For simplicity, we use a periodic weight update in the following part of this work.

To allow a meaningful examination of the influence of time-varying hardware impairments, we model such effects by a single parameter per effect. The variations are thus simulated by a simple *random walk* model for this sole parameter. The effects we consider are:

1) IQ-imbalance: we use a simple model only depending on a single parameter $\beta_{\text{IQ}} \in \mathbb{R}$ as

$$y = \beta_{\text{IQ}} \cdot \mathcal{R}(x) + (1 - \beta_{\text{IQ}}) \cdot j \cdot \mathcal{I}(x) \qquad (2)$$

2) Non-linearity and clipping: we model this effect by a single parameter $\gamma_{\text{NL}} \in \mathbb{R}$ describing AM-AM distortion by a third-order non-linear function with normalized input as given in [14]

$$g(x) = x - \gamma_{\text{NL}} |x|^2 x \qquad (3)$$

[3]Besides convolutional codes, any other channel coding scheme such as low-density parity-check (LDPC) codes and Polar codes could be applied straightforwardly. We opted for convolutional codes as it provides a flexible codeword length. To avoid high simulation complexity in combination with the Python/Tensorflow setup, we limit $\mu = 2$, however, further gains are expected with increasing $\mu$.

[4]Remark: $\tilde{\mathbf{x}}$ and $\hat{\mathbf{x}}$ do not need to be stored necessarily, as they can be calculated during the forward-propagation of the SGD anyway.

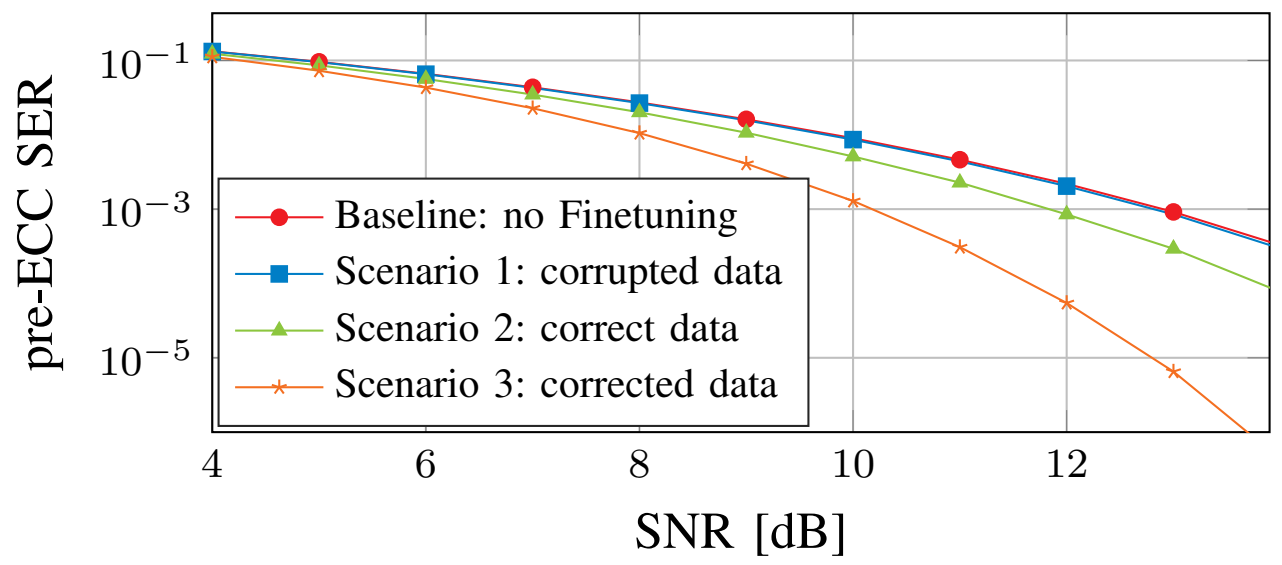

Fig. 3: Impact of corrupted training data during finetuning of a trainable OFDM communications system suffering from TX IQ-imbalance. All setups use the same amount of 20,000 OFDM symbols as training data, collected at SNR of 5 dB, which equals a pre-ECC SER of 9.6%.

with $x \in \mathbb{C}$ and $|x|$ in $[0, 1]$, i.e., we clip $x$ such that $|x| \leq 1$ and keep the phase of $x$ unchanged.

Furthermore, the channel adds additive white Gaussian noise (AWGN).

### A. Training with corrupted labels

We first analyze the impact of corrupt labels during training for IQ-imbalance. Thus, we analyze three different scenarios:

1) Training set contains corrupted data, i.e., transmission errors are not detected/corrected
2) Training set contains only the correctly received symbols, i.e., transmission errors are detected but not corrected (e.g., due to a cyclic redundancy check (CRC))
3) All symbols in the training set are correct, i.e., transmission errors are corrected (ECC)

To avoid side effects, all training data sets have the same size of 20 batches with 1,000 OFDM symbols per batch. Fig. 3 depicts the same finetuning step (from $\beta_{\text{IQ}} = 0.45$ to $\beta_{\text{IQ}} = 0.65$) for the listed scenarios and shows that finetuning based on corrupted labels does not significantly improve the *pre-ECC* performance of the trainable OFDM system. It also shows that finetuning based only on correctly received symbols helps improving the performance slightly, but as somehow expected and shown in Fig. 3, the performance increases significantly if corrected (via ECC) symbols are used as labels during the finetuning process. In other words *from errors one learns*.

## IV. Adaptivity

In the following, we show that the proposed finetuning can be used to compensate impairments on-the-fly and without the need of explicit labels. The underlying channel parameters $\beta_{\text{IQ}}$ and $\gamma_{\text{NL}}$ are provided in Fig. 4 and Fig. 5, respectively. The horizontal red lines indicate that a finetuning step was triggered and the weights are updated. As the considered effects are typically slowly changing, it seems to be reasonable to assume that an unlimited amount of training data is available. During each time step $\Delta t$, $N_{\Delta t}$ OFDM symbols are transmitted and used for the finetuning procedure. The system performs 7 finetuning SGD iterations per time step using all $N_{\Delta t}$ OFDM symbols.

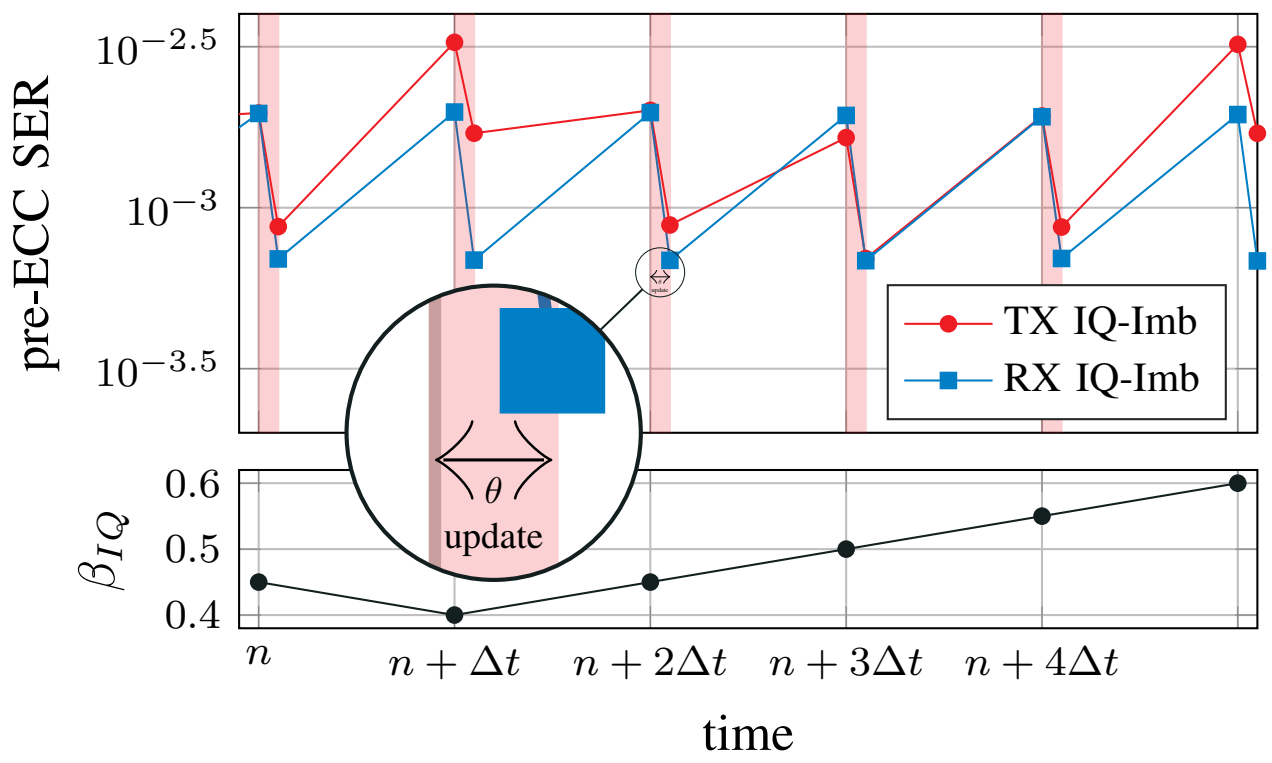

Fig. 4: SER performance at SNR of 10 dB for varying IQ-imbalance $\beta_{\text{IQ}}$. The horizontal red areas indicate receiver finetuning solely based on labels recovered at the receiver. $N_{\Delta t} = 5,000$ OFDM symbols are transmitted and used for training per time step.

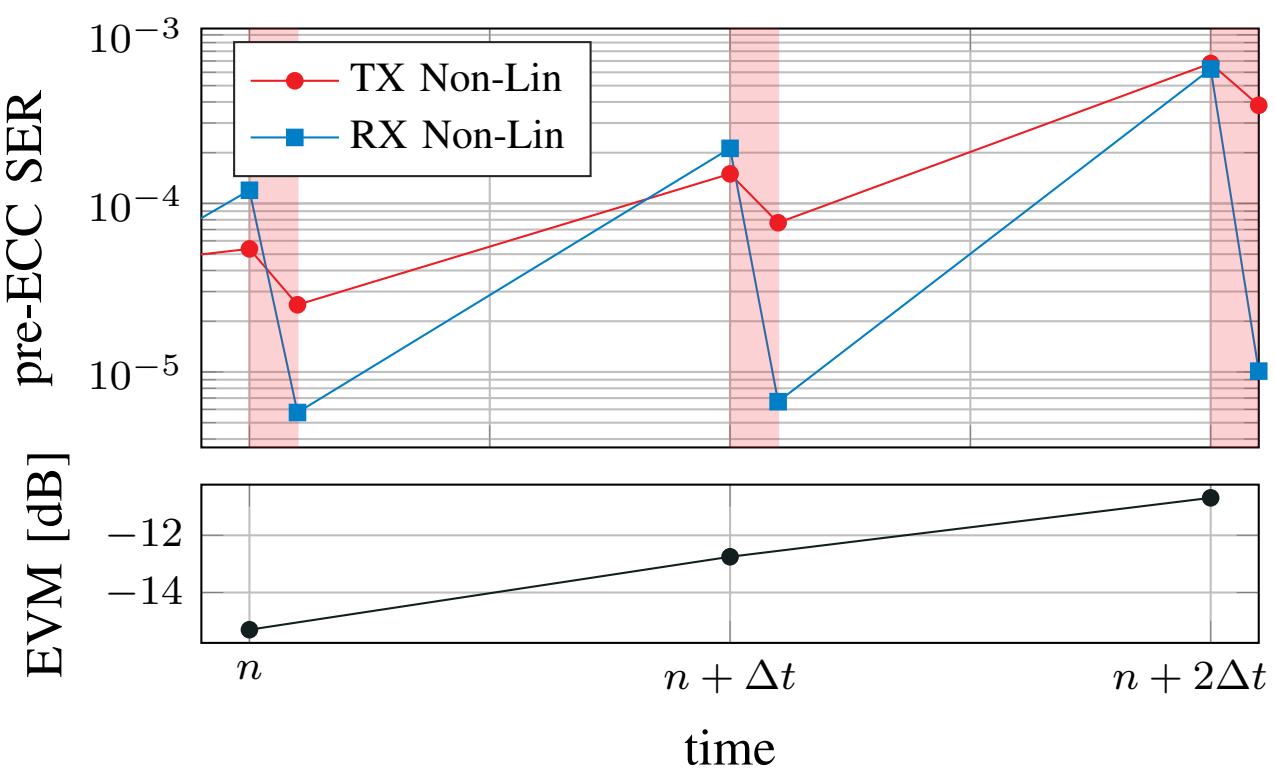

Fig. 5: SER performance at SNR of 14 dB for varying non-linearity $\gamma$. The horizontal red areas indicate receiver finetuning solely based on labels recovered at the receiver. $N_{\Delta t} = 20,000$ OFDM symbols are transmitted and used for training per time step.

### A. IQ-Imbalance

In this experiment, we study the impact of IQ-imbalance as given in (2) for transmitter and receiver side, respectively. The system is originally trained for $\beta_{\text{IQ}} = 0.5$, however, we assume that $\beta_{\text{IQ}}$ now changes during time as shown in Fig. 4. In Fig. 4, it can be seen that both effects, transmitter and receiver IQ-imbalance, can be compensated with the residual network by only using the received data. As expected, the transmitter is slightly harder to compensate, but generally tolerates IQ-imbalance. Keep in mind that only the receiver weights are updated during the finetuning.

### B. Non-linearity

In the next experiment, we examine the impact of a non-linearity and clipping as given in (3) (no IQ-imbalance assumed), again at transmitter and receiver side. As the transmitter does not change during finetuning, we can provide the error vector magnitude (EVM) caused by the non-linearity (see [13]) as shown in Fig. 5. The non-linearity turns out to be much harder to compensate when compared to IQ-imbalance, which could be explained by the fact that the non-linearity and the clipping effect destroy the transmitted information irreversibly rather than just cause a distortion. However, finetuning can still improve the receiver performance significantly as shown in Fig. 5. Again, we observe that the transmitter non-linearity is more difficult to compensate than the receiver side non-linearity.

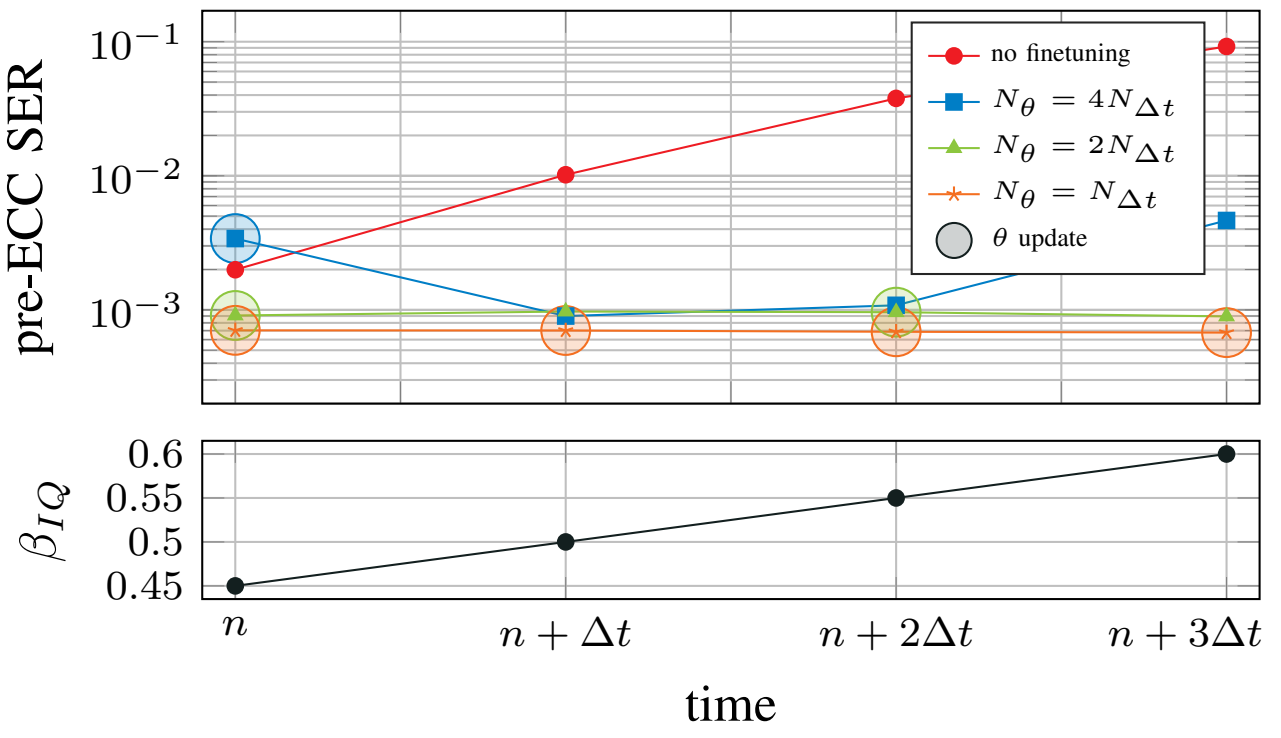

Fig. 6: SER performance at SNR of 10 dB for posterior finetuning for varying IQ-imbalance.

### *C. Posterior Finetuning*

We now analyze the impact of outdated weights during the decoding. This could be practically motivated whenever the coherence time of the channel is below the training time required for finetuning. Therefore, we assume that during a time step $\Delta t$ an amount of $N_{\Delta t}$ OFDM symbols is transmitted with constant channel conditions. A practical motivation of this scenario may be the downlink in deep space communication where the signal is fully recorded at the receiver and the ground station may have (almost) infinite resources to recover the signal without further latency constraints. Due to outdated weights, or in other words when the mismatch between trained and actual channel conditions is too large, the receiver operates at high BER which cannot be compensated by the ECC (see red curve Fig. 6). Thus, the finetuning gains are degraded due to spoiled labels (see Fig. 3).

To overcome this problem, we propose *posterior finetuning*[5] in such a way that intermediate weight updates are only based on a subset of $N_\theta \leq N_{\Delta t}$ OFDM symbols. The detailed algorithm is given in Algorithm 1, where $N_\theta$ denotes the number of OFDM symbols used for the SGD update (*window-size*). As can be seen in Fig. 6 for the effect of IQ-imbalance, a proper choice of $N_\theta$ (in relation to $N_{\Delta t}$) can improve the decoding performance significantly.

**Algorithm 1:** Posterior Finetuning

**Input :** Number of OFDM-symb. per SGD update $N_\theta$
Recorded sequence of $m \cdot N_\theta$ OFDM-symbols
$\mathbf{Y} = [\mathbf{y}_0, \dots, \mathbf{y}_{m \cdot N_\theta - 1}]$
Pre-trained pre-equalizer weights $\boldsymbol{\theta}$

**Output:** Estimated information bit sequence
$\hat{\mathbf{U}} = [\hat{\mathbf{u}}_0, \dots, \hat{\mathbf{u}}_{m \cdot N_\theta - 1}]$

/* Sliding window decoding */

**for** $\tau \leftarrow 0$ **to** $m-1$ **do**

  /* Receive with given weights $\theta$ */

  $[\hat{\mathbf{u}}_{\tau \cdot N_\theta}, \dots, \hat{\mathbf{u}}_{(\tau+1)\cdot N_\theta - 1}] \leftarrow \texttt{decode}(\boldsymbol{\theta}; [\mathbf{y}_{\tau \cdot N_\theta}, \dots, \mathbf{y}_{(\tau+1)\cdot N_\theta - 1}])$

  /* Re-encode */

  $[\hat{\mathbf{x}}_{\tau \cdot N_\theta}, \dots, \hat{\mathbf{x}}_{(\tau+1)\cdot N_\theta - 1}] \leftarrow \texttt{re-encode}([\hat{\mathbf{u}}_{\tau \cdot N_\theta}, \dots, \hat{\mathbf{u}}_{(\tau+1)\cdot N_\theta - 1}])$

  /* Update weights $\theta$ */

  $\boldsymbol{\theta} \leftarrow \texttt{sgd}([\mathbf{y}_{\tau \cdot N_\theta}, \dots, \mathbf{y}_{(\tau+1)\cdot N_\theta - 1}]; [\hat{\mathbf{x}}_{\tau \cdot N_\theta}, \dots, \hat{\mathbf{x}}_{(\tau+1)\cdot N_\theta - 1}])$

**end**

## V. Conclusions and Outlook

We have examined the possibilities of mitigating the influence of corrupted labels for finetuning of the trainable components by using an ECC and, thus, to recover a labeled training set on the receiver side without knowing the transmitted sequence, i.e., without the need for pilots. This is due to the fact that we deal with man-made signals and, thus, can influence the structure of the transmitted data by adding redundancy. We have shown that the proposed method can compensate unforeseen impairments without being explicitly considered in the original design of the system. This has been exemplified by inducing IQ-imbalance and non-linearity in both transmitter and receiver. As expected, effects occurring at the transmitter side are much harder to compensate than receiver side effects.

[5]In case the required time for finetuning $t_{ft} > \Delta t$, posterior decoding may be based on a recorded sequence and, thus, the name posterior.

## References

[1] A. Caciularu and D. Burshtein, "Blind channel equalization using variational autoencoders," *arXiv preprint arXiv:1803.01526*, 2018.

[2] D. Neumann, T. Wiese, and W. Utschick, "Learning the MMSE channel estimator," *IEEE Transactions on Signal Processing*, 2018.

[3] E. Nachmani, E. Marciano, D. Burshtein, and Y. Be'ery, "RNN decoding of linear block codes," *arXiv preprint arXiv:1702.07560*, 2017.

[4] T. Gruber, S. Cammerer, J. Hoydis, and S. ten Brink, "On deep learning-based channel decoding," in *Proc. of CISS*, 2017.

[5] N. Farsad and A. Goldsmith, "Detection algorithms for communication systems using deep learning," *arXiv preprint arXiv:1705.08044*, 2017.

[6] N. Samuel, T. Diskin, and A. Wiesel, "Deep MIMO detection," *SPAWC*, pp. 690–694, 2017.

[7] E. Nachmani, Y. Be'ery, and D. Burshtein, "Learning to decode linear codes using deep learning," *54th Annu. Allerton Conf. Commun., Control, Comput. (Allerton)*, pp. 341–346, 2016.

[8] T. J. O'Shea, K. Karra, and T. C. Clancy, "Learning to communicate: Channel auto-encoders, domain specific regularizers, and attention," *IEEE Int. Symp. Signal Process. Inform. Tech. (ISSPIT)*, pp. 223–228, 2016.

[9] S. Dörner, S. Cammerer, J. Hoydis, and S. ten Brink, "Deep learning-based communication over the air," *IEEE Journal of Selected Topics in Signal Processing*, 2018.

[10] I. Goodfellow, Y. Bengio, and A. Courville, *Deep Learning*. MIT Press, 2016.

[11] K. He, X. Zhang, S. Ren, and J. Sun, "Deep residual learning for image recognition," in *Proceedings of the IEEE conference on computer vision and pattern recognition*, 2016, pp. 770–778.

[12] T. J. O'Shea and J. Hoydis, "An introduction to machine learning communications systems," *IEEE Transactions on Cognitive Communications and Networking*, vol. 3, no. 4, pp. 563–575, Dec 2017.

[13] A. Felix, S. Cammerer, S. Dörner, J. Hoydis, and S. ten Brink, "OFDM-Autoencoder for End-to-End Learning of Communications Systems," *ArXiv e-prints*.

[14] T. Schenk, *RF imperfections in high-rate wireless systems: impact and digital compensation*. Springer Science & Business Media, 2008.